\documentclass{appolb}
\usepackage{epsfig}
\newcommand{\Journal}[4]{{#1}{\bf #2}\ (#3)\ #4}
\newcommand{\AP}{Ann.\ Phys.\ }
\newcommand{\ARNS}{Ann.\ Rev. \ Nucl. \ Sci.\ }

\newcommand{\IJMP}{Int.\ J.\ Mod.\ Phys.\ }
\newcommand{\JP}{J.\ Phys.\ }

\newcommand{\NCA}{Nuovo\ Cimento\ {\bf A}}
\newcommand{\NCL}{Nuovo\ Cimento\ Lett.\ }
\newcommand{\NIM}{Nucl.\ Instr.\ and Meth.\ }

\newcommand{\NPB}{Nucl.\ Phys.\ {\bf B}}

\newcommand{\PL}{Phys.\ Lett.\ }
\newcommand{\PLB}{Phys.\ Lett.\ {\bf B}}
\newcommand{\PR}{Phys.\ Rev.\ }

\newcommand{\PRL}{Phys.\ Rev.\ Lett.\ }

\newcommand{\PRD}{Phys.\ Rev.\ {\bf D}}

\newcommand{\RepPP}{Rep.\ Prog.\ Phys.\ }
\newcommand{\RMP}{Rev.\ Mod.\ Phys.\ }
\newcommand{\SJNP}{Sov.\ J.\ Nucl.\ Phys.\ }
\newcommand{\SP}{Sov.\ Phys.\ -\ JETP}
\newcommand{\ZP}{Z.\ Phys.\ }
\newcommand{\ZPA}{Z.\ Phys.\ {\bf A}}

\newcommand{\pptopp}{\bf{\it {\cal PP}2{\cal PP}}}
\newcommand{\sigtot}{\sigma_{\rm tot}}

\newcommand{\dsigel}{d\sigma_{\rm el}/dt}
\newcommand{\pomeron}{I \! \!P}

\newcommand{\lumi}{{\cal L}}

\newcommand{\pbarp}{{\it \overline{p}p}}
\newcommand{\roots}{\sqrt{s}}
\newcommand{\gevsq}{${\rm GeV}^{2}$\ }
\newcommand{\gevsqd}{${\rm GeV}^{2}.$\ }
\begin{document}
\title{The $\pptopp$ Experiment at RHIC%
\thanks{Presented at Cracow Epiphany Conference on Quarks and Gluons,
3 -- 6 January, Cracow, Poland}%
}
\author{J. Chwastowski
\address{Henryk Niewodnicza\'nski Institute  of Nuclear Physics,\\ 
Cracow\\[10pt]
(for The $\pptopp$ Collaboration  \cite{collab})}
}\maketitle
\begin{abstract}
The $\pptopp$ experiment is devoted to the proton -- proton elastic scattering 
measurements at the Relativistic Heavy Ion Collider (RHIC) at the 
centre-of-mass energies 
$50  \leq \roots \leq 500 $ GeV and the four-momentum transfer
$0.0004 \leq |t| \leq 1.3$ \gevsqd The option of polarized proton beams offers 
a unique possibility to investigate the spin dependence of the proton
elastic scattering in a systematic way.
 The energy dependence 
of the total and elastic cross section, the ratio of the real to the imaginary 
part of the forward scattering amplitude, and the nuclear slope parameter
will be studied.
In the medium $|t|$ region ($|t| \leq 1.3$ \gevsq) the energy dependence of 
the dip structure in the elastic differential cross section will be measured.
With polarized beams the measurement of spin dependent observables:
the difference of the total cross sections as function of the of initial 
transverse spin states, the analyzing power and the double spin asymmetries
will be used to map the $s$ and $t$ dependence of the proton helicity 
amplitudes.
\end{abstract}
  
\section{Introduction}
The elastic channel contributes about 20\% to the total cross section.
This, coupled with the necessity of explaining diffraction in terms of QCD,
makes the nucleon -- nucleon elastic scattering a very attractive tool for the 
exchange mechanism investigation.
So far, the elastic proton -- proton scattering was studied up to the highest 
ISR energies ($\roots = 62$ GeV) while for $\pbarp$ the measurements
reached $\roots = 1.8$ TeV. The ISR data confirmed earlier prediction 
\cite{heisenberg,regge} of the total cross section rise with energy. Also 
Pomeranchuk's prediction \cite{pomeranchuk} on the asymptotic behaviour of the
total cross sections for $pp$ and $\pbarp$ scattering was extensively tested. 
Summary of elastic scattering measurements and phenomenological models can be 
found in \cite{elsumm}.

With advent of
the Relativistic Heavy Ion Collider (RHIC) at Brookhaven National Laboratory 
a new energy domain opens for a study of the proton - proton scattering.
The accelerator is capable of delivering proton beams \cite{rhicspin} polarized
up to 70\% with luminosities reaching $\lumi = 10^{31}\rm{cm}^{-2}\rm{s}^{-1}$.
This offers a unique opportunity for a systematic study of  
the spin dependent and 
spin independent elastic scattering in a new, mostly unexplored centre-of-mass
energy range of $50  \leq \roots \leq 500 $ GeV and for the four-momentum 
transfer $0.0004 \leq |t| \leq 1.3$ \gevsq \cite{proposal}.

The colliding of polarized beams allows for the measurement of the 
proton helicity structure in wide range of $|t|$ and energy 
within a single experiment. The polarization studies will shed a light onto 
emerging new picture of diffraction and its spin dependence \cite{buttimore}.

\section{Physics Programme}

\subsection{Unpolarized Scattering}

The QCD proved to be very successful in describing hadronic interactions at 
large $|t|$ where perturbative methods are applicable. However, for small and 
medium $|t|$ values in elastic and diffractive scattering, in a the long-range
domain, the QCD has not yet provided accurate predictions. This regime of soft 
hadronic interactions is a field of phenomenological models constrained with 
asymptotic theorems. The Regge-Gribov theory \cite{regge} is based on the 
analycity, unitarity and crossing symmetry of the scattering amplitude and 
describes the interaction via exchange of trajectories related to the poles 
(or cuts) in the complex angular momentum plane. The elastic channel is treated
as a shadow of many inelastic channels. The elastic amplitude is considered to 
be mainly imaginary and helicity conserving. Phenomenological fits 
\cite{dl92,cu00,dl98,bl99} based on Regge theory are able to describe all the 
hadronic and photoproduction total cross sections in the full energy range. 
The data are well described by the exchange of the Pomeron ($\pomeron$) and  
lower-lying Regge meson, the reggon, trajectories. The Pomeron is described in 
the QCD language  \cite{qcdpom} in terms of multi-gluon exchange and point-like
coupling to quarks: a color singlet two-gluon exchange with $C = +1$ 
corresponds to  the Pomeron while the color singlet three-gluon $C = -1$ state 
corresponds to the Odderon \cite{odderon}. In spite of its successes, the 
Regge theory still lacks a firm explanation on the QCD grounds. High precision 
measurements in the energy and $|t|$-range covered by $\pptopp$ are certain to
 have a large impact on this data-driven field.

\begin{figure}[ht]
\epsfysize=15.5cm
\centerline{\epsffile{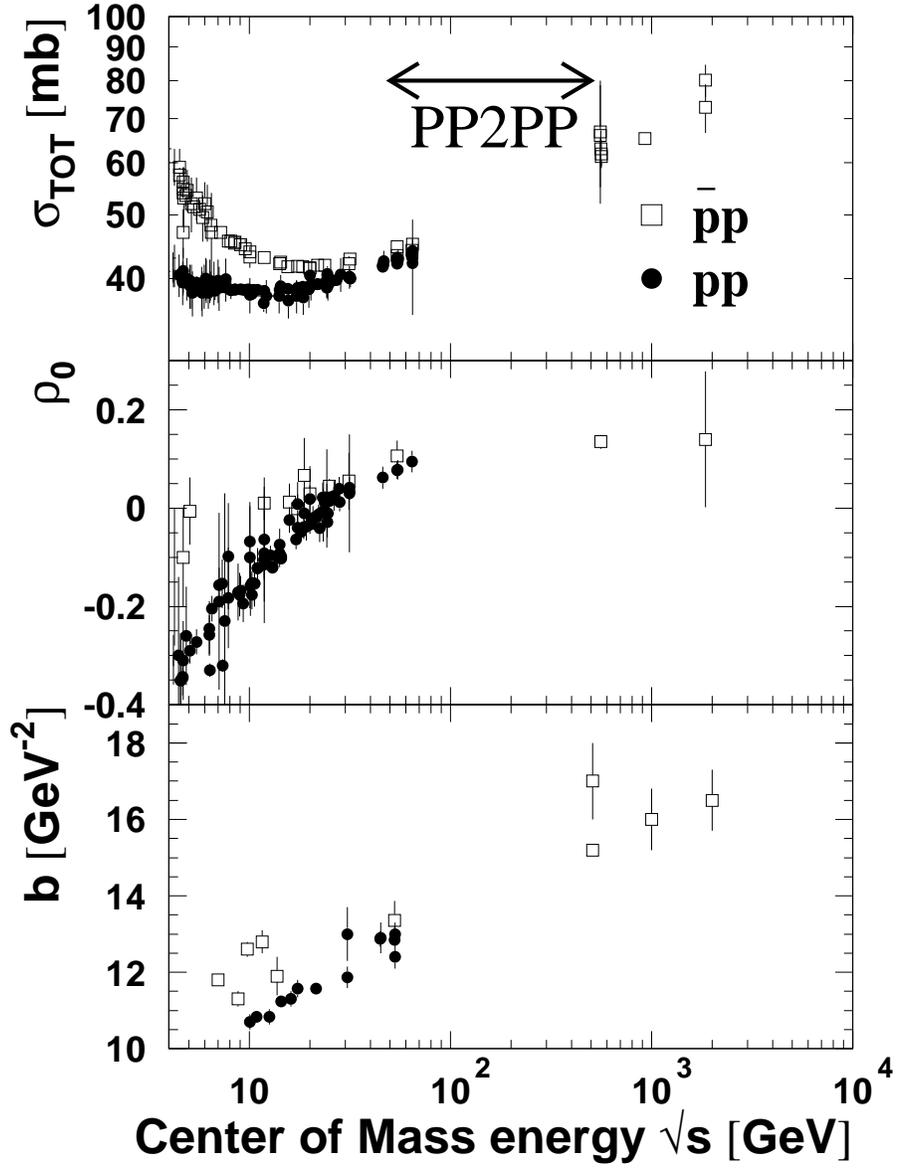}}
\caption{The total cross section, $\sigtot$, the $\rho_{0}$ parameter and the 
nuclear slope, $b$, as a function of the centre-of-mass energy.}
\label{fig:sigtot}
\end{figure}

Figure \ref{fig:sigtot} summarizes present measurements of the total cross 
section $\sigtot$, the ratio of real to imaginary part of the forward amplitude
$\rho_{0}$, and the nuclear slope $b$, for both $pp$ and $\pbarp$ interactions.
The energy range of the $\pptopp$ experiment is also shown. It partially
overlaps the high energy ISR region and reaches the low end of the  
$S\overline{p}pS$ range. A comparison with existing data at the same energies 
will be advantageous.

\begin{figure}[ht]
\epsfxsize=12cm
\centerline{\epsffile{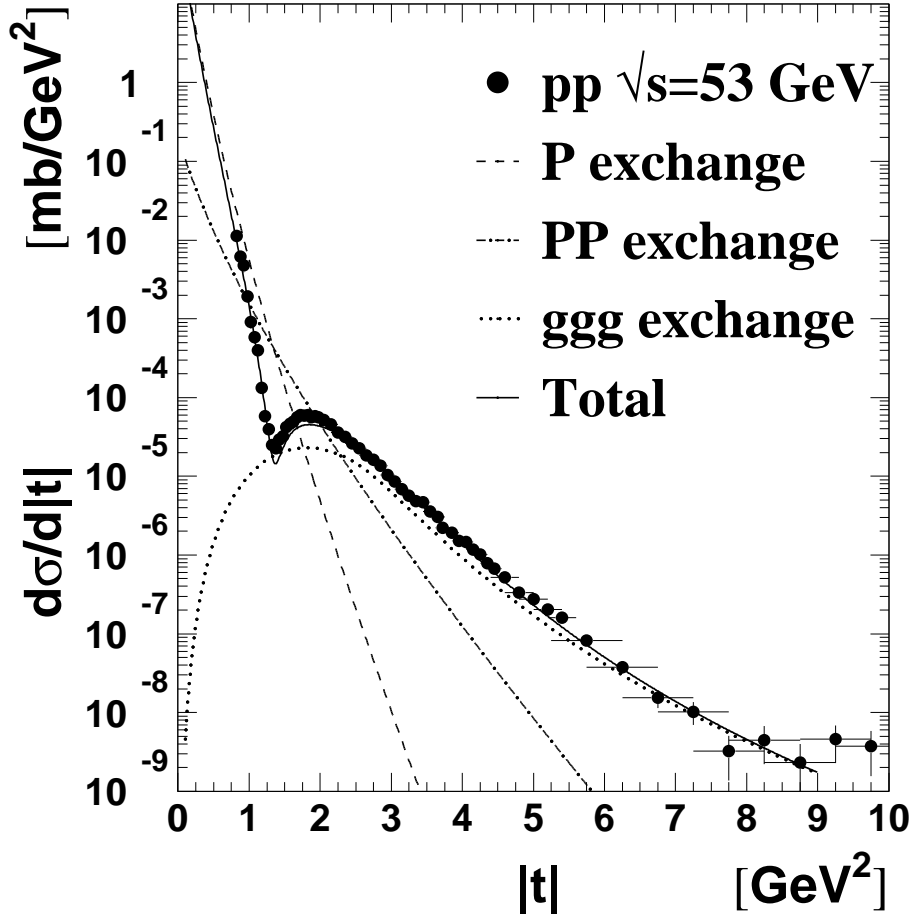}}
\caption{The differential cross section, in $pp \rightarrow pp$ at 
$\roots = 53$ GeV 
\cite{sigteldata}. The curves are the model \cite{dl84} predictions.}
\label{fig:sigel}
\end{figure}
\noindent
The differential elastic cross section $\dsigel$ as measured at ISR  
\cite{sigteldata} is shown in Fig. \ref{fig:sigel} together with the
phenomenological model \cite{dl84} predictions. The data show characteristic 
features: a forward peak followed by an exponential fall, a local minimum above
$ -t = 1$ \gevsq and a secondary maximum or a broad shoulder. The position of 
the minimum moves toward smaller $|t|$ values with increasing energy. A similar
trend is observed for $\pbarp$ data. However, the depth of the minimum is 
smaller than that seen for $pp$. The data are well described 
by the model predictions.

The $|t|$ coverage of $\pptopp$ includes the ``diffraction dip'' region. This 
local minimum was explained \cite{dl84} by the interference of the exchange 
amplitudes. In particular, the three-gluon exchange gives a negative 
contribution in case of $pp$ interaction while for $\pbarp$ scattering it is 
positive.

In the small $|t|$ region the $\pptopp$ experiment covers also the Coulomb 
amplitude dominated domain. Since the Coulomb amplitude is absolutely known 
then a very small $|t|$ measurement will give an absolute measurement of the 
accelerator luminosity and an absolute normalization of the hadronic amplitude.

\subsection{Polarized Scattering}

The understanding of the spin dependence of the scattering amplitudes at high 
energies and small $|t|$ is very important. It challenges the QCD since it 
involves its applications  in the long-range, non-perturbative regime. New 
measurements coming from RHIC will test the QCD prediction at a new level of 
accuracy and detail.

The $pp$ polarization data are commonly discussed using the s-channel helicity 
amplitudes $\phi_{i} (i = 1,...,5$) \cite{helamp}.
Table \ref{tab:ampli} lists the linear combinations of $\phi_{i}$ with definite
quantum numbers exchanged asymptotically \cite{lincomb}.
\begin{table}[ht]
\centering
\caption{s-channel amplitudes.}
\vspace*{0.1cm}
\label{tab:ampli}
\begin{tabular}{l|c}
\hline
Naturality & Helicity \\
\hline
N$_{0}$& $(\phi_{1}+\phi_{3})/2$\\ 
N$_{1}$& $\phi_{5}$\\ 
N$_{2}$& $(\phi_{4}-\phi_{2})/2$\\ 
U$_{0}$& $(\phi_{1}-\phi_{3})/2$\\ 
U$_{1}$& $(\phi_{4}+\phi_{2})/2$\\ 
\hline
\end{tabular}
\end{table}

The N and U amplitudes correspond to natural and unnatural parity exchange and 
the subscripts 0, 1  and 2 describe the total s-channel helicity flip involved.
The spin-averaged cross sections are given by:
$$
\sigtot = \frac{1}{K}Im(N_{0}(0))
$$
and
$$
\frac{d\sigma}{dt} = \frac{1}{16\pi{K^{2}}}(|N_{0}|^{2}+2|N_{1}|^{2}+|N_{2}|^{2}
                                            +|U|_{0}|^{2}+|U_{2}|^{2})
$$
where $ K = \sqrt{s(s-4m^{2})}$. The most commonly used spin-dependent 
observables are listed below. The indices stand for different spin 
orientations: $N$ - along the normal to the scattering plane, $S$ - along the 
transverse direction in the scattering plane, and $L$ - along the particle 
direction. The differences of the total cross section for the anti-parallel and
parallel spin orientations are
$$
\Delta\sigma_{N} = \sigma{\uparrow\downarrow}-\sigma{\uparrow\uparrow} = 
		\frac{1}{K}Im(N_{2}(0)-U_{2}(0))
$$
$$
\Delta\sigma_{L} = \frac{1}{K}Im(U_{0}(0)).
$$
The analysing power,  the single-spin asymmetry, is
$$
{\cal A}_{N} \frac{d\sigma}{dt} = -2\frac{1}{16\pi{K^{2}}}Im\left\{(N_{0}-N_{2})N_{1}^{\star}\right\},
$$
and the double-spin  asymmetries are defined as
$$
{\cal A}_{NN} \frac{d\sigma}{dt} = -2\frac{1}{16\pi{K^{2}}}
	\left\{Re(U_{0}U_{2}^{\star}-N_{0}N_{2}^{\star})+|N_{1}|^{2}\right\},
$$
$$
{\cal A}_{LL} \frac{d\sigma}{dt} = -2\frac{1}{16\pi{K^{2}}}
	\left\{Re(N_{2}U_{2}^{\star}-N_{0}U_{0}^{\star})\right\},
$$
$$
{\cal A}_{SS} \frac{d\sigma}{dt} = -2\frac{1}{16\pi{K^{2}}}
	\left\{Re(N_{0}U_{2}^{\star}-N_{2}U_{0}^{\star})\right\}.
$$
To achieve a full amplitude analysis one needs a substantial number of 
measurements in the same kinematic region. 
Up till now, the polarization experiments were carried out for $\roots$ up to 
28 GeV in fixed target environment. This severely affected the minimum values 
of accessible $|t|$ with exception of the E581/704 Collaboration which used 
``live-target'' to reach down to the CNI region \cite{e704}. It is worth 
pointing out that these problems are non-existing in the RHIC environment since
the measurements are performed for protons freely moving in vacuum.

Existing data \cite{asymdata} on the analysing power at 
$|t| \approx 0.1 - 2.5$ \gevsq show that 
\begin{itemize} 
\item at $|t| < 0.4 $ \gevsq ${\cal A}_{N}$  decreases like $1/s$ up to 
$s \approx 50$ \gevsq and possibly flattens-off around small positive value,
\item  ${\cal A}_{N}$  becomes negative for $ 0.4 < |t| < 1 $ \gevsq and 
$s \geq 50$ 
\gevsq, reaches a minimum and again increases crossing zero in the 
``diffraction dip'' region.
\item for large $|t|$ the analysing power is constant or possibly decreases 
to zero above $t \geq 2.5$ \gevsqd
\end{itemize}
The Regge theory predicts \cite{qcdpom,15} that
the ${\cal A}_{N}$ should decrease as $s^{-\alpha}$, with $0.5 < \alpha < 1$, 
since at 
lower energies hadronic spin-flip amplitude should be dominated by the reggon 
exchanges. This is confirmed by the data which in the  beam momentum range of 
45 -- 300 GeV/c show very little energy dependence. However, small values of 
the analysing power can be compatible with zero. On the other hand, the 
negative values of  ${\cal A}_{N}$ for $ 0.4 < |t| < 1 $ \gevsq cannot be 
deduced from the standard Regge picture.

\begin{figure}[ht]
\epsfxsize=12cm
\centerline{\epsffile{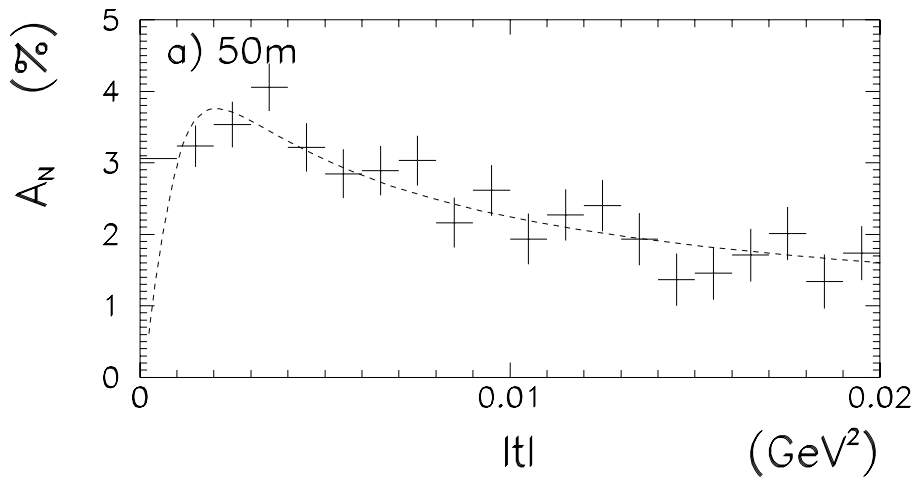}}
\caption{The theoretical analysing power ${\cal A}_{N}$ (dashed line) 
superimposed with 
simulated measurements at $\roots = 500$ GeV for stations at 50 m.}
\label{fig:anacc}
\end{figure}

A characteristic shape of the ${\cal A}_{N}(t)$ was predicted \cite{19} in the
CNI region. The analysing power increases from zero, reaches a maximum at 
$|t| \approx 0.003$ \gevsq and slowly decreases with increasing $|t|$. This 
structure results from the interference of the electromagnetic spin-flip and 
hadronic non-flip amplitudes. The maximum ${\cal A}_{N}$ is about 5\% and 
shows a very mild energy dependence. $\pptopp$, in its first physics run will 
cover the region around the ${\cal A}_{N}$ maximum and the simulations show 
that the analysing power can be precisely measured. Figure \ref{fig:anacc} 
shows the theoretical predictions \cite{22} with simulated data from the 
detectors placed 50 meters away from interaction region (IR).

\begin{figure}[ht]
\epsfxsize=11cm
\centerline{\epsffile{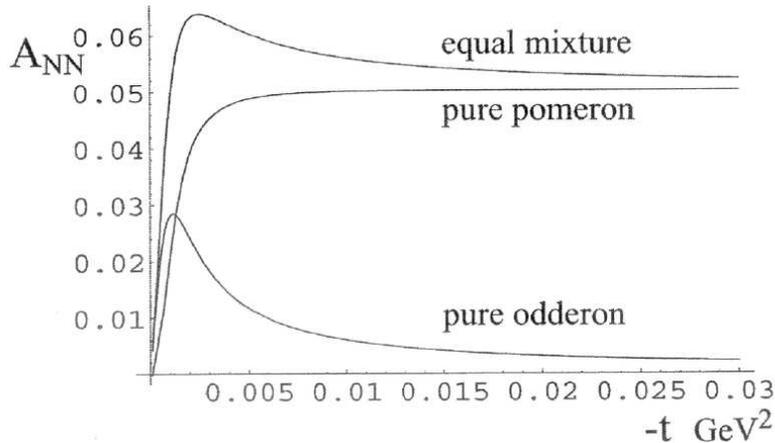}}
\caption{The Odderon and Pomeron contributions to the  double spin asymmetry 
${\cal A}_{NN}$ from \cite{18}.}
\label{fig:annacc}
\end{figure}
With both RHIC beams polarized, the double spin asymmetry can be measured 
in a wide $|t|$ range. In the small $|t|$ region it will help to answer the 
question of the Odderon contribution. A spin-flip amplitude $\phi_{5}$ does not
necessarily decreases with increasing energy. This can be due to various 
mechanisms like peripheral pion production, a di-quark component in the proton,
instanton effects or due to the Odderon contribution \cite{17}.
In Fig. \ref{fig:annacc}, the theoretical predictions \cite{18} for the 
Pomeron and Odderon contributions to ${\cal A}_{NN}$  are shown. 
In the higher $|t|$-region further investigation of experimental observation 
of large differences for between the cross sections for parallel and 
anti-parallel spin orientations will be performed. The experiment \cite{27} 
shows that the protons appear to interact harder when their spins are aligned. 
However, it is not clear whether this effectis present for higher $\roots$.
The measurement of the spin dependent observables in the high-$|t|$ region 
will reveal the helicity structure of the exchanges. It will address the 
questions of non-vanishing  spin-flip amplitudes in diffraction, the Odderon 
contribution or whether hard scattering produces large spin effects.
They give an insight into the details of the long-range QCD interaction between
the protons and possibly how the perturbative regime is approached.
 
\section{The Experimental Setup}

The measurement of the small angles observed in elastic scattering requires 
interfacing of the detectors to the accelerator lattice.

\subsection{Method}

The proton scattering takes place at the interaction region (IR) at a point
($x^{\star}, y^{\star}$), where  $x^{\star}(y^{\star})$ is the horizontal 
(vertical) coordinate in the plane perpendicular to the accelerator axis.
Later the particle traverses inside the beam pipe and is, eventually, 
registered at (x,y) by the detector. Since the accelerator parameters are know
then the scattering angle -- $\Theta^{\star}_{y}$ and the position -- 
$y^{\star}$ at the IR can be calculated from the measured ones.\\ 
If $\Psi$ denotes the phase advance and $\beta$ is the betatron function at 
the detection point then 

$$ y = \sqrt{\frac{\beta}{\beta^{\star}}}
        (\cos{\Psi}+\alpha^{\star}\sin{\Psi}){y^{\star}}
	+ \sqrt{{\beta}{\beta^{\star}}}\sin{\Psi}~{\Theta^{\star}_{y}}
$$

where $\beta^{\star}$ is the betatron function at the IR, 
$\alpha^{\star}$ is its derivative and $\Theta^{\star}_{y}$ is the proton 
scattering angle. For the purpose of this 
experiment the machine lattice configurations with $\alpha^{\star} = 0$ were 
considered.\\
The above formula can be re-written as
\begin{equation}
y = a_{11}{y^{\star}} + {\cal L}_{eff}{\Theta^{\star}_{y}}
\label{equ:parallel1}
\end{equation}
with 
$$
a_{11} =\sqrt{\frac{\beta}{\beta^{\star}}}
        (\cos{\Psi}+\alpha^{\star}\sin{\Psi})
$$
and
$$
{\cal L}_{eff} = \sqrt{{\beta}{\beta^{\star}}}\sin{\Psi}.
$$
The  optimum experimental conditions are for $a_{11} = 0$ $(\alpha^{\star} = 0,
\Psi = (2n+1)\pi/2)$  and ${\cal L}_{eff}$
as large as possible. Then eq. \ref{equ:parallel1} becomes
\begin{equation}
y =  {\cal L}_{eff}{\Theta^{\star}_{y}}.
\label{equ:parallel}
\end{equation}
This means that the scattering angle can be determined from the position 
measurement alone. In other words the particles scattered at the same 
angles at the IR will be focused onto the same point. This conditions is 
known as the ``parallel-to-point'' focusing. 

Another concern for the experimental set-up optimization is the minimum value
of the four-momentum transfer $|t|$, namely $t_{min}$. This value depends on 
the distance of the detectors to the proton beam. It should be as small as 
possible to reach the Coulomb dominated region, yet large enough so the machine
beams are not destroyed. The value of $t_{min}$ is determined by the minimum 
scattering angle, ${\Theta_{min}}$, by
$$
 t_{min} = (p\Theta_{min})^{2},
$$
where p is the proton absolute momentum and
$$
\Theta_{min} = \frac{d_{min}}{{\cal L}_{eff}}, \hspace{0.5cm} d_{min}= k{\sigma_{y}}+d_{0}
$$
where $d_{min}$ is the minimum distance of the approach to the beam,
k is the machine dependent factor (10--20) which can be optimized via the beam 
scraping, $\sigma_{y}$ denotes the vertical beam size at the detection point 
and $d_{0}$ measures the detector ``dead-space''. If $d_{0}$ is small then

$$
t_{min} = \frac{\epsilon}{\beta}{k^{2}}{p^{2}}
$$
which means that the the smallest $t_{min}$ can be reached making the betatron
 function as large as possible and reducing the $k$-factor and the beam 
emittance, $\epsilon$.

It is intended to measure the elastic and diffractive scattering for
small and medium $|t|$ intervals.

\subsection{Layout of the Experiment}

\begin{figure}[ht]
\epsfysize=5cm
\epsfxsize=12cm
\centerline{\epsffile{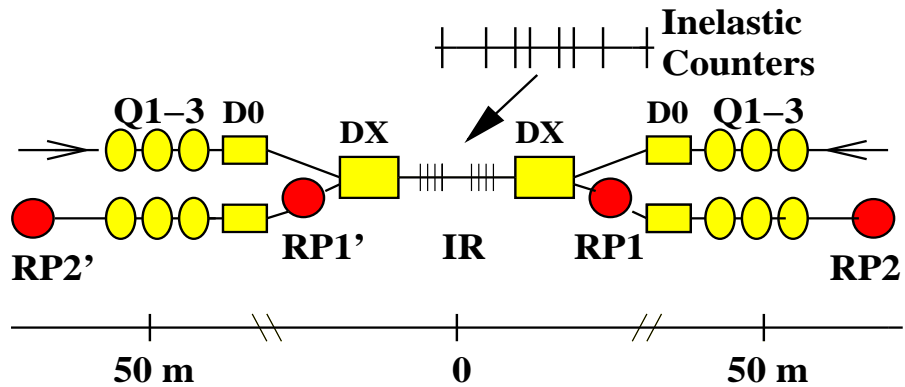}}
\caption{Layout of the $\pptopp$ experiment.}
\label{fig:layout}
\end{figure}

Protons scattered at small angles are registered with detectors positioned in 
the immediate vicinity of the beam. This is achieved with help of a Roman
Pot (RP) station \cite{sigteldata,ua4} placed at the positions for which 
condition \ref{equ:parallel} is fulfilled. A layout of the experiment is shown
in Fig. \ref{fig:layout}. In the drawing magnetic elements of the machine 
lattice ($DX$, $D0$ and $Q1-Q3$) are shown; symbol $RP1,{2}$ denotes the 
Roman Pot station.

\begin{figure}[ht]
\epsfysize=8cm
\epsfxsize=8cm
\centerline{\epsffile{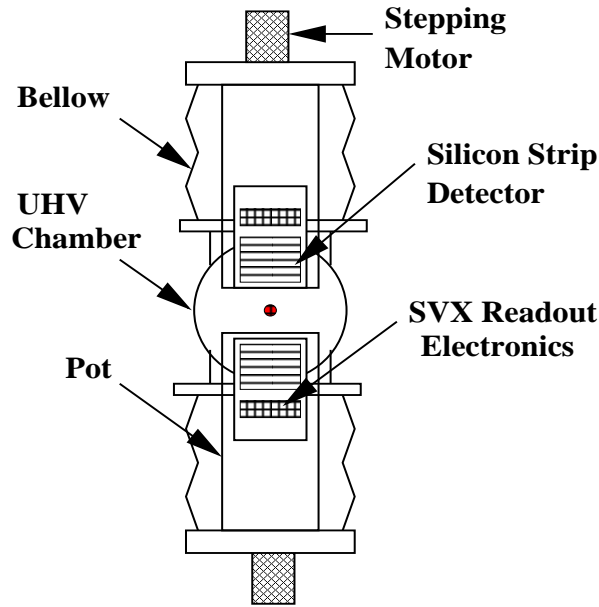}}
\caption{A schematic view of the Roman Pot station.}
\label{fig:station}
\end{figure}

The pot, see  Fig. \ref{fig:station}, allows to insert the detector into
the accelerator beam pipe. The active part of the detector consists of four 
planes of  silicon strip detectors out of which two planes measure vertical 
and two horizontal coordinates. 
The detectors, 8 by 5 cm$^{2}$, with 100 $\mu$m strip-to-strip and 450 $\mu$m
strip-to-edge spacing, are read by the readout system based on the D0 SVX-IIe
128-channel front-end chip 
and the Stand-Alone Sequencer 
.
This type of detector proved to be a reliable and stable device. It provides:
\begin{itemize}
\item a very good uniformity of efficiency,
\item small dead area between the detector active part end the beam,
\item many highly efficient detector layers,
\item small cells,
\item good spatial resolution.
\end{itemize}

\subsection{Trigger}

There are two basic trigger configurations. The elastic trigger is derived from
the scintillator counters placed behind the silicon detectors. The second, 
inelastic, is used to trigger on diffractive or inelastic events and consists 
of eight scintillator counters placed symmetrically around the interaction 
region. The trigger requests a coincidence of the beam crossing and 
the appropriate scintillators' signals.

\begin{figure}[ht]
\epsfxsize=13cm
\epsfxsize=13cm
\centerline{\epsffile{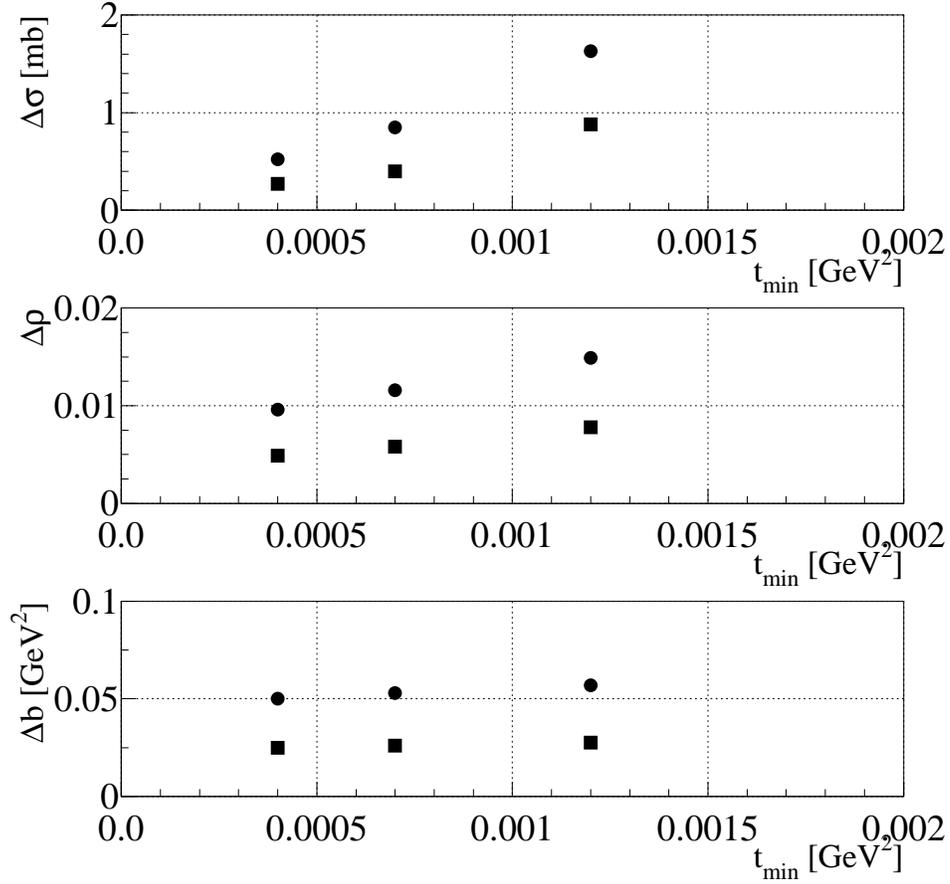}}
\caption{The uncertainty on the total cross section $\Delta\sigma$, the ratio
real to imaginary part of the forward amplitude $\rho_{0}$, and the nuclear
slope $b$ as a function of the minimum accessible $|t|$, for 0.8 million 
events (full dots) and 3.2 million events(full squares) statistics.}
\label{fig:syst}
\end{figure}
\subsection{Measurements at Small $|t|$}

A special tune of the RHIC accelerator allows reaching the Coulomb region, at
$|t|$ values down to 0.0004 GeV$^{2}$. This tune will allow the measurements 
of very small angles (down to 10 $\mu$rad). Since this tune is not compatible 
with high luminosity running in other interaction regions it will require a 
dedicated running period and is postponed beyond 2005.
\begin{table}[ht]
\centering
\caption{Parameters in the Monte Carlo simulation.}
\vspace*{0.1cm}
\label{tab:mcparam}
\begin{tabular}{l|c}
\hline
Parameter & Value \\
\hline
Beam momentum        & 250 GeV/c\\ 
Beam momentum spread & 250 MeV/c\\ 
Crossing angle of beams $\Theta^{0}_{x,y}$  & 5 $\mu$rad,5 $\mu$rad\\ 
Error on crossing angle $\Delta\Theta^{0}_{x,y}$  & 6 $\mu$rad \\ 
Detector offsets     & 20 $\mu$m\\
Detector resolution in x,y     & 100 $\mu$m\\
Interaction region size in z:$\sigma_{z}$ & 15 cm\\
Beam emittance: $\epsilon$& 5$\pi$ mm mrad\\
\hline
\end{tabular}
\end{table}
The detector acceptance covers CNI region and reaches well into the Coulomb 
amplitude dominated domain hence it allows for a precise determination of
$\sigma_{tot}$, $\rho_{0}$ and $b$ parameters. To evaluate the detector 
performance a Monte Carlo was used. Table \ref{tab:mcparam} lists major 
parameters and their values used in the simulation.

The angular distribution of protons was simulated with small angle CN cross 
section formula parameterized with $\sigma_{tot}$, $\rho_{0}$ and $b$. 
Using the transport equation the position of the scattered 
protons at the detector was predicted. Such simulated data were input to the 
fitting program yielding $\sigma_{tot}$, $\rho_{0}$ and $b$.
Resulting uncertainties on the parameters are shown as a function of $t_{min}$
in Fig. \ref{fig:syst}. Two different statistics, $0.8\cdot 10^{6}$ and 
$3.2\cdot 10^{6}$, equivalent to one day of stable data taking are compared.

The major contributions to the uncertainties are due to the statistics and
the  $t_{min}$ cutoff.

\subsection{Measurements at Medium $|t|$}

\begin{figure}[ht]
\epsfysize=14cm
\centerline{\epsffile{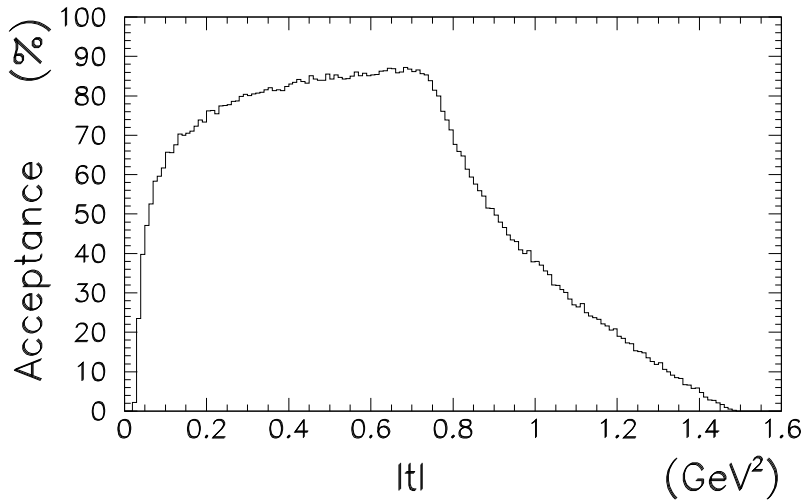}}
\vspace{-5cm}
\caption{The medium-$|t|$ acceptance of the detectors.}
\label{fig:medtacc}
\end{figure}

The collinearity requirement on the out-going protons will provide 
identification of the small angle elastic scattering events. Table 
\ref{tab:rpexpcomp} compares parameters of two earlier \cite{sigteldata,ua4} 
and $\pptopp$ experiments.
\begin{table}[hb]
\caption{Parameters of the ISR, UA4 and $\pptopp$ experiments.}
\vspace*{0.1cm}
\label{tab:rpexpcomp}
\begin{tabular}{l|c|c|c}
\hline
Parameter &  ISR & UA4 & $\pptopp$ \\
\hline
CMS energy (GeV)& 23-62& 546-630 & 50-500\\
Luminosity ${\cal L} (cm^{-2}s^{-1}$& few 10$^{30}$ & few 10$^{28}$ & few 
10$^{31}$ \\
Maximum $|t|$ (\gevsq) & 10 & $\approx$2& $\approx$1.3\\
Momentum resolution $\Delta p/p$&$\approx$5\% & $\approx$0.6\% & 
$\approx$1.5\%\\
$|t|$  resolution $\Delta t$&$\approx 0.015\sqrt{|t|}$&
$\approx 0.06\sqrt{|t|}$&$\approx 0.02\sqrt{|t|}$\\
\hline
\end{tabular}
\end{table}
For the RHIC tune  $\beta^{\star} = 10$m, the normalized
emittance is $\epsilon = 20{\pi}$mm mrad, and at $\roots = 500$ GeV the size 
and the angular spread of the beam at the IR are $\sigma_{y} = 0.45$ mm
and $\sigma_{\Theta y} = 45$ ${\mu}$rad, respectively. These parameters were
used to calculate the medium-$|t|$ acceptance  shown in Fig. \ref{fig:medtacc}.
The acceptance is larger than 20\% for $0.03 \leq |t| \leq 1.3$ \gevsqd This is
the interval which covers the ``diffractive dip'' region at $\roots = 500$ GeV.
At larger $|t|$ values the acceptance is limited due to the DX magnet's 
aperture cut at 5.4 mrad. At $|t| = 1$ \gevsq, the expected cross section is 
0.001 mb/\gevsqd The detector acceptance is about 35\%. At RHIC luminosity 
${\cal L} = 10^{31} cm^{-2}s^{-1}$ about 10$^{4}$ elastic events per $|t|$-bin
of 0.05 \gevsq per day is expected which shows that running time is not a 
limitation.

\section{Running Plan}

The data collection will be staged due to the experimental setup staging and
the accelerator operation mode requirements.

In 2002, the experimental apparatus will consists of the two fully equipped 
stations. These stations will be placed in the RHIC tunnel about 57 meters away
from the interaction region. In January, it is planned to collect the data at 
$\roots = 200$ GeV. The accelerator will operate with  $\beta^{\star} = 10$m. 
The useful range of the four-momentum transfer will be: 
$0.004 \leq -t \leq 0.02$ \gevsqd With collected data it will be possible to 
measure the nuclear slope parameter and possibly the single spin asymmetry 
${\cal A}_{N}$.

In 2003, there will be four fully equipped stations used for the data taking.
It is planned to run at $\roots = 200$ GeV for $0.0004 \leq -t \leq 0.02$ 
\gevsqd This run will be devoted to the study of the CNI region and the energy 
dependence of $\sigtot$, $\rho$, $b$ and spin asymmetries: ${\cal A}_{N}$, 
${\cal A}_{NN}$. About one day of stable running is necessary to
obtain the error on analysing power of $\Delta A_{N} \approx 0.2-0.3\%$ for 
$A_{N}$ value $\approx 4\%$.

For 2004, a major update of the experimental setup is foreseen. Additional 
Roman Pot stations will be placed in the straight sections connecting the $DX$ 
and $D0$ magnets, see Fig. \ref{fig:layout}. This will enable the measurements 
for $-t \leq 1.3$ \gevsqd The energy dependence of the nuclear slope, elastic 
differential cross section and spin observables will be investigated. Also the
 diffractive minimum region will be studied. At luminosity 
$\lumi = 4\cdot 10^{30} {\rm cm}^{-2}{\rm s}^{-1}$ about 200 hours of the data 
taking will be needed to acquire 1000 events per 0.02
\gevsq bin around $-t = 1$\gevsqd

Investigation of the very small $|t|$ region will require a special, $\pptopp$ 
dedicated tune of the accelerator, $\beta^{\star} \approx 100$m. Therefore 
precision measurements of ratio of the real-to-imaginary part of the forward 
amplitude, of the total cross section for $0.0004 \leq -t \leq 0.12$ \gevsq 
will be postponed until after 2004.

It is worth noting that the same experimental setup can be used to measure the 
elastic scattering of $pd$, $p^{\uparrow}d$, $dd$ and $p^{\uparrow}~^{4}He$.
It is also possible to extend the four-momentum region beyond 1.3 \gevsq
by adding RP stations in front of the $DX$ magnets.

\section{Summary}

The $\pptopp$ experiment at RHIC Collider has been introduced. The physics 
programme includes a systematic study of the spin dependent and spin 
independent features of the elastic proton -- proton scattering in wide and 
mostly unexplored energy domain $50  \leq \roots \leq 500 $ GeV and for the
four-momentum transfer $0.0004 \leq |t| \leq 1.3$ \gevsq within a single 
experiment. The use of the same apparatus will play a crucial role in 
understanding and reduction of systematic errors. Elastically scattered protons
will be measured with silicon detectors placed in the Roman Pots. New and 
precise measurements coming from the $\pptopp$ experiment will help to test 
the QCD in a non-perturbative regime. They will also shed light on the helicity
structure of the proton and that of the object exchanged in the interaction.


\end{document}